\documentstyle[epsf]{elsart}

\begin{document}
\begin{flushleft}
{{\tt
DESY 95-170 \hfill ISSN 0418-9833\\
ITP-UH 23/95\\
hep-ph/9509313
}}
\end{flushleft}
\vspace{1cm}

\begin{frontmatter}

\title{Improved hard-thermal-loop effective action\\
                 for hot QED and QCD}

\thanks[DFG]{This work is partially supported by Deutsche
Forschungsgemeinschaft (DFG) under grant no.~Schu 1045/1-1.}

\author{Fritjof Flechsig\thanksref{email}}
\address{Institut f\"ur Theoretische Physik,
         Universit\"at Hannover, \\
         Appelstr. 2, D-30167 Hannover, Germany}

\and

\author{Anton K. Rebhan\thanksref{addra}}

\thanks[email]{e-mail: flechsig@itp.uni-hannover.de}
\thanks[addra]{Address after November 1, 1995:
        Institut f\"ur Theoretische Physik,
        Technische Universit\"at Wien,
        Wiedner Hauptstr. 8--10,
        A-1040 Vienna, Austria;\\
        e-mail: rebhana@email.tuwien.ac.at}

\address{DESY, Gruppe Theorie,\\
        Notkestra\ss e 85, D-22603 Hamburg, Germany}

\begin{abstract}

The conventional results for hard thermal loops, which are
the building blocks of resummed perturbation theory in
thermal field theories, have collinear singularities when
external momenta are light-like. It is shown that by taking
into account asymptotic thermal masses these singularities
are removed. The thus improved hard thermal loops can be
summarized by compact gauge-invariant effective actions,
generalizing the ones found by Taylor and Wong, and by
Braaten and Pisarski.

\end{abstract}
\end{frontmatter}
\newpage

%%%%%%%%%%%%%%%%%%%%%%%%%%%%%%%%%%%%%%%%%%%%%%%%%%%%%%%%%%%%
%%%%%%%%
%%%%%%%%  a few macros
%%%%%%%%

\newcommand{\bea}{\begin{eqnarray}}
\newcommand{\beal}[1]{\begin{eqnarray}\label{#1}}
\newcommand{\eea}{\end{eqnarray}}
\newcommand{\beq}{\begin{equation}}
\newcommand{\eeq}{\end{equation}}
\newcommand{\bel}[1]{\begin{equation}\label{#1}}
\newcommand{\nn}{\nonumber}

\newcommand{\Sp}{\,\mbox{tr}\,}

\newcommand{\sumint}
  {{\textstyle\sum}\!\!\!\!\!\!\int\nolimits_{\;P}}
\newcommand{\sumints}
  {{\scriptstyle\Sigma}\!\!\!\!\int\nolimits_{P}}

\newcommand{\w}{\omega}
\renewcommand{\a}{\alpha}
\renewcommand{\b}{\beta}
\renewcommand{\d}{\delta}
\newcommand{\s}{\sigma}
\renewcommand{\L}{\Lambda}
\newcommand{\G}{\Gamma}
\newcommand{\g}{\gamma}
\newcommand{\bg}{{\mbox{\protect\boldmath $\gamma$}}}

\newcommand{\mn}{{\mu\nu}}

\newcommand{\cl}[1]{{\cal #1}}

\newcommand{\0}{\over }
\newcommand{\1}[1]{\frac{1}{#1}}
\newcommand{\2}{\frac{1}{2}}
\newcommand{\4}{\frac{1}{4}}
\newcommand{\6}{\partial}
\newcommand{\with}{\quad\mbox{with}\quad}
\newcommand{\quer}[1]{\overline{#1}}

\def\({\left(}     \def\){\right)}
\def\lek{\left[}   \def\rek{\right]}
\def\lgk{\left\{}  \def\rgk{\right\}}
\newcommand{\rang}{\right\rangle}
\newcommand{\lang}{\left\langle}

\def\lesssim{\mbox{\,\raisebox{.3ex}{
           $<$}$\!\!\!\!\!$\raisebox{-.9ex}{$\sim$}\,\,}}
\def\gtrsim{\mbox{\,\raisebox{.3ex}{
           $>$}$\!\!\!\!\!$\raisebox{-.9ex}{$\sim$}\,\,}}

\def\vc#1{{\bf{#1}}}
\def\wu#1{\sqrt{{#1} \,}^{ \hbox to0.2pt{\hss$
        \vrule height 2pt width 0.6pt depth 0pt $} \;\! } }

%%%%%%%%%%%%%%%%%%%%%%%%%%%%%%%%%%%%%%%%%%%%%%%%%%%%%%%%%%%%
%%%%%%%%
%%%%%%%%  Maintext
%%%%%%%%

\section{Introduction}

It is a typical phenomenon in thermal field theory that the
conventional perturbative loop expansion breaks down because
of infrared singularities which are usually cured by the
generation of thermal masses\cite{Kapu}. In the particularly
simple case of a massless $\lambda\phi^4$-theory, the
leading temperature contribution to the self-energy consists
of a simple mass term with $m=\wu\lambda T$. An improved
perturbation theory requires that this thermal mass be
treated on a par with the tree-level inverse propagator
whenever the momentum of the latter becomes comparable to
(or smaller than) the former. A consistent way of doing so
is to add the thermal mass term to the bare Lagrangian and
to subtract it again through counterterms at higher loop
orders.

In general, however, the thermal self-energy is more
complicated than a constant mass squared and it may also be
the case that higher vertex functions are equally important.
This is the case in QED and QCD, for which a resummed
perturbation theory has been developed by Braaten and
Pisarski\cite{BP}. The leading temperature contributions to
be resummed have been obtained by Frenkel and
Taylor\cite{FT} and independently by the former, who have
coined the term ``hard thermal loops''. These hard thermal
loops (HTL) turned out to satisfy ghost-free Ward
identities\cite{BPward,KKR}. Manifestly gauge invariant as
well as rather compact expressions for the effective
Lagrangians which summarize them were found in
Refs.~\cite{eff,shortcut,BPeff,Nair}.

This resummation program has been applied successfully to a
number of problems (see e.g.~the forthcoming review in
Ref.~\cite{appl}). Occasionally, however, it turned out that
the thermal masses contained in the HTL effective action
were insufficient to screen all infrared singularities. For
example, through quasi-particle mass-shell singularities,
the damping rates of plasma excitations with nonzero
momentum as well as the next-to-leading order screening
corrections are blown up by unscreened magnetostatic modes
\cite{FRS}. However, assuming the existence of some
effective infrared cutoff at the (unfortunately entirely
nonperturbative) magnetic mass scale at least allows one to
obtain a leading logarithmic correction.

Another shortcoming of the conventional HTL resummation has
been revealed recently in the attempt to calculate the
production rate of soft real photons from a quark-gluon
plasma. In the case of hard real photons, a complete leading
order calculation was carried out in Ref.~\cite{BNNR}, where
it was found that the mass singularities of the quarks are
screened by HTL corrections. But with soft real photons, the
hard thermal loops themselves introduced uncancelled
collinear singularities \cite{BPS}. We shall concentrate on
this latter type of singularities in what follows.

In a quite different context a similar difficulty was
encountered in Ref.~\cite{KRS}. In the simpler gauge theory
of scalar electrodynamics, the complete next-to-leading
order dispersion laws of photonic excitations could be
obtained in a resummed one-loop calculation. However, the
perturbative result for the longitudinal branch turned out
to become unreliable as the light cone was approached with
increasing plasmon momentum. The reason for this breakdown
was again a collinear singularity of a HTL diagram. There it
was found that it can be removed by a further resummation.

In this paper, we extend this strategy to the case of QED
and QCD and we shall show that a manifestly gauge invariant
HTL effective action can be found which is completely free
from collinear singularities. It improves upon the effective
action of Refs.~\cite{eff,shortcut,BPeff} with which it
coincides for external momenta that are sufficiently far
from the light-cone.

In the next section we argue that the hard propagators have
to be dressed by asymptotic thermal masses whenever
collinear singularities make the hard thermal loops
themselves sensitive to such higher-order corrections. In
sect.~3 we carry out a resummation of the asymptotic thermal
masses first for the case of purely gluonic QCD, and in
sect.~4 when fermions are included. The analytic structures
obtained differ, but in each case the improved hard thermal
loops can be summarized by a gauge invariant effective
action. Sect.~5 contains our conclusions and gives an
outlook to potential applications.

\section{Hard thermal loops in the vicinity of the light-cone}

Consider the 00-component of the polarization tensor
(i.e. self-energy) at leading order as given by its
hard thermal loop. It reads universally \cite{KalKlW}
\bea
\Pi_{00}(Q) &=& 4e^2 \int\!{d^3p \0 (2\pi)^3}\,n(p)
\lgk 1-{Q_0 \0
 Q_0-\vc p\vc q/p} \rgk \nn\\
&=& 3m^2  \(1 - {Q_0 \0 2q} \ln{Q_0+q\0Q_0-q}\)
\label{poohtl}\eea
with $m=eT/3$ and $n$ is the Bose function, where $e$
is either the coupling constant of QED or
\bel{edef}
e=g\wu{N+N_f/2}
\eeq
for SU($N$) with $N_f$ fermions.

Close to the light-cone $\Pi_{00}$ is
found to diverge logarithmically,
\beq
\Pi_{00} \sim -m^2\ln\1{\varepsilon}
\with \varepsilon^2=Q^2/q^2 \; .
\label{lndiv}\eeq
The origin of this divergence is a collinear singularity in
the loop integral (\ref{poohtl}) at
$\vc p\vc q=p q$ when $Q_0=q$.

The next-to-leading order contribution for soft momenta
$Q\equiv(Q_0,\vc q)$, $Q_0,\vc q \sim eT$ is given by resummed
one-loop diagrams. In both scalar electrodynamics \cite{KRS}
and QCD \cite{FS}, it turns out to diverge like
\beq
\d\Pi_{00} \sim m^2\frac{e}{\varepsilon}
\label{div}\eeq
when approaching the light-cone.
Evidently, the resummed perturbative series breaks down for
$\varepsilon\lesssim e$, for then $\d\Pi_{00} \gtrsim \Pi_{00}$.

In the case of scalar electrodynamics (massless and without
scalar self-inter\-act\-ions) it is relatively easy to
analyse this problem \cite{KRS}. The collinear singularity
in (\ref{lndiv}) is brought about by the masslessness of the
internal scalar particles. The even stronger singularity in
(\ref{div}) is generated by a premature restriction to soft
loop momenta. The scalar propagators are dressed and
therefore massive, but one has to subtract off the
contributions already covered by the bare HTL diagram. This
again involves massless scalar propagators, which become
large at the light-cone irrespective of the momentum scale.
Thus (\ref{div}) reveals a latent UV-divergence.

\begin{figure}
\begin{center}
\unitlength0.7mm
\parbox{28mm}{
\begin{picture}(40,10)
  \put(2,5){\line(1,0){36}}
  \put(20,5){\circle*{5}}
\end{picture}}
=
\parbox{21mm}{
\begin{picture}(30,10)
  \put(2,5){\line(1,0){26}}
\end{picture}}
+
\parbox{42mm}{
\begin{picture}(60,24)
  \put(2,12){\line(1,0){13}}
  \put(35,12){\line(1,0){25}}
  \put(25,12){\circle{20}}
  \put(25,2){\circle*{4}}
  \put(25,22){\circle*{4}}
  \put(35,12){\circle*{5}}
  \put(47,12){\circle*{5}}
\end{picture}}
\parbox{12cm}
{\caption{Prototypical Schwinger-Dyson equation --- for
simplicity for a $\phi^3$-theory. Internal propagators are
always dressed, whereas vertices appear in bare and dressed
versions.\label{fig_sd}}
}
\end{center}
\vspace{1cm}
\end{figure}
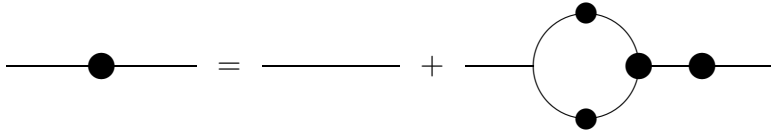

However, from the Schwinger-Dyson equations for the various
Green functions (Fig.~\ref{fig_sd}) it is clear that in the
full theory the scalar lines are always massive with mass
squared $m^2=\4 e^2 T^2(1+O(e))$. The collinear
singularities of $\Pi_{00}$ at the light-cone, which are
produced by scalar particles, should therefore be spurious.

Indeed, keeping the thermal masses also for the hard scalar
propagators yields a finite result for the HTL contribution
to $\Pi_{00}$ where $\varepsilon$ in (\ref{lndiv}) gets
replaced by $e$. The increasingly singular contributions
(\ref{lndiv},\ref{div}) are thus seen to get cancelled by
higher loop diagrams corresponding to a hard scalar loop
where scalar self-energy diagrams are inserted repeatedly.
Away from the light-cone, such insertions are suppressed by
powers of $e^2$, but they cease to be so when
$\varepsilon\lesssim e$.

Resumming thermal masses already at the stage where the hard
thermal loops are being calculated raises the question about
the systematics of this procedure. For this it is useful to
think in terms of a renormalization group approach
\cite{BrRG}. The conventional resummation program of Braaten
and Pisarski can be understood as a two-step procedure.
First an arbitrary scale $\L$ with $T\gg\L\gg eT$ is
introduced which divides all momenta and energies in hard
and soft, and an effective theory is built from integrating
out all the hard modes. To leading order, the result is
independent of the actual value of $\L$, and is given by the
HTL effective action. In a second step, Green functions with
external soft momenta are calculated using this effective
action with soft loop momenta. The degree to which a
perturbative expansion based on the effective theory can
possibly make sense is limited by the accurateness of the
effective action. Higher loop orders of the former require
sufficiently high loop orders of the latter.

In the above example it became apparent that for soft
momenta with $\varepsilon\lesssim e$ the effective action
receives contributions from hard diagrams with arbitrarily
high loop orders. In order to sum them systematically one
would have to solve the Schwinger-Dyson equations, which are
an infinite set of coupled equations. However, the present
singularities are caused primarily by the masslessness of
the bare propagators, whereas the structure of the
Schwinger-Dyson equations is such that only full propagators
enter (in contrast to vertex functions which appear in both
bare and dressed forms). As long as it is sufficient to sum
only self-energy insertions to produce massive propagators,
there is no particular problem with overcounting.

In fact, also in the more complicated gauge theories like
spinor electrodynamics and QCD the physical degrees of
freedom retain thermal masses for momenta $p\gg eT$. In the
case of gauge bosons, there is a transverse and a
longitudinal branch of quasi-particle mass-shells, but the
longitudinal branch rapidly dies out with increasing
momentum. In the HTL approximation, the longitudinal
plasmons approach the light-cone, but in doing so the
corresponding residue vanishes exponentially \cite{Pispha}.
The transverse mode, on the other hand, tends toward an
effective asymptotic mass
\beq
  m_\infty^2\equiv\Pi_t(Q^2=0) = \frac{e^2T^2}{6}+O(e^2T\L) \;
\label{mu}
\eeq
and residue 1. (Recall that in QCD $e$ is defined by
(\ref{edef}).) Likewise, the dispersion laws for
ultrarelativistic fermions have a collective branch (the
``plasmino'' with a flipped relation between helicity and
chirality) which dies out with increasing momentum and a
normal one that remains, again with an asymptotic mass
proportional to $e^2T^2+O(e^2T\L)$ (without spoiling
chirality) \cite{KliW}.

The HTL effective action is generated entirely by the
physical degrees of freedom of the bare theory, which is
particularly evident when the Coulomb gauge is used, where
only transverse modes are heated.\footnote{By employing a
nonstandard real-time formalism, this feature can be kept
also for covariant gauges \cite{LR}.} To leading order,
these contribute only as real particles living on their
mass-shell, which is the light-cone $Q^2=0$.
Correspondingly, the collinear singularities present in
conventional HTL diagrams with external light-like momenta
are all removed when the above asymptotic thermal masses for
transverse gauge bosons and any other ultrarelativistic
particles are included.

It is an important feature of the asymptotic thermal masses
that the result (\ref{mu}) as well as the corresponding one
for fermions (see below) does not depend on the condition
$Q_0, q\ll T$ which is otherwise essential for the
derivation of the HTL results. It holds for arbitrary
momentum as shown in Eq.~(A.21) of Ref.~\cite{KrKrR}.
Moreover, the definition (\ref{mu}) is itself not sensitive
to a soft modification of the hard propagators, because
$\Pi_t$ is not singular at the light-cone.

However, the smallness of the asymptotic masses might help
formerly negligible contributions of higher-order diagrams
to increase when approaching the light-cone. This may raise
the necessity to reorganize the perturbation series.

In gauge theories one might expect that giving masses to the
transverse gauge boson modes without modifying also their
vertices might violate gauge invariance. Nevertheless, we
shall see presently that gauge invariance of the HTL
effective action is maintained if the above procedure is
followed to generalize hard thermal loops to soft lightlike
momenta.

\section{Purely gluonic QCD}

In this section we shall first treat purely gluonic QCD,
where $e^2=g^2N$. The improved HTL diagrams are obtained
almost exactly as usually, the only difference being that
the gluon propagator for hard momenta $p\ge\L$ is modified
to take into account the asymptotic thermal mass for
transverse gluons,
\beq
G_\mn(P) = A_\mn \Delta_m + (B_\mn + \a D_\mn)\Delta_0
\label{Gmn}
\eeq
\bea
&&{\rm with} \qquad
\Delta_m = \1{P^2-m_\infty^2} \quad,\quad
\Delta_0 = \1{P^2}\nn\\
&&{\rm and} \qquad
A_\mn=g_\mn-B_\mn-D_\mn \quad,\quad
B_\mn={V_\mu V_\nu\0 V^2} \quad , \quad
D_\mn={P_\mu P_\nu\0 P^2}
\eea
where $V_\mu= U_\mu P^2-P_\mu(PU)$ and $U_\mu=(1,\vc 0)$ is
the four velocity of the thermal bath at rest.

\subsection{Gluon self-energy}

Since the collinear singularities in the bare one-loop
polarization tensor are only logarithmic, the extraction of
the $T^2$ contribution with soft external momenta can
proceed as usually (see, however, the Appendix for a
pitfall). Soft external momenta in the numerator can be
neglected when compared to hard loop momenta, and the
familiar expression is obtained, but with $\Delta_m$ in
place of the originally massless propagators.

\begin{figure}
     \centerline{ \epsfxsize=4in
     \epsfbox[55 395 370 500]{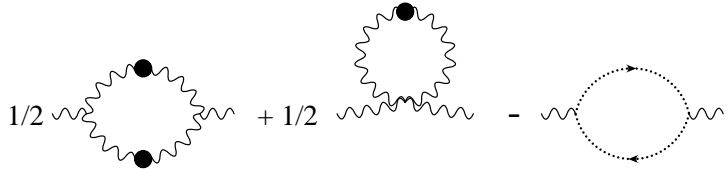} }
 \begin{center}
  \parbox{12cm}
  {\caption{The contributions to the improved gluon
  self-energy. The wavy lines with a blob represent gluon
  propagators whose transverse piece resums the asymptotic
  thermal mass; the dotted line represents the (unchanged)
  ghost propagator.\label{fig_gse}} }
 \end{center}
\vspace{1cm}
\end{figure}

Let us demonstrate this in the example of $\Pi_{00}(Q)$.
With the modified gluon propagator (\ref{Gmn}), the diagrams
shown in Fig.~\ref{fig_gse} respectively yield
\beal{piooltg}
\Pi_{00}^{\rm loop}(Q)  & = &
  g^2N\sumint\lgk \Delta_m - 4\Delta_m^-\Delta_m P_0 K_0
   - \Delta_0^-\Delta_0 P_0 K_0 \rgk \nn\\
\Pi_{00}^{\rm tad}(Q)   & = &
 -g^2N\sumint 3\Delta_m \quad \\
\Pi_{00}^{\rm ghost}(Q) & = &
 g^2N\sumint \Delta_0^-\Delta_0 P_0 K_0 \; ,\nn
\eea
such that in the sum only massive propagators appear,
\beq
\Pi_{00}(Q) = g^2N \sumint \lgk - 2\Delta_m
   - 4\Delta_m^-\Delta_m P_0K_0  \rgk \; .
\label{lc50}
\eeq

In the following we use the imaginary-time formalism, so the
zero components of the momenta $P=(P_0,\vc p)$ are discrete
Matsubara frequencies $P_0=2\pi i nT$. The symbol $\sumints$
is defined as
\beq
\sumint = T\sum_n\int {d^3p\0 (2\pi)^3}
\eeq
Throughout our paper $Q$ is the external momentum, $P$ is
summed over and $K$ is the difference $K=Q-P$. An index
${}^-$ means the transformation $P\to K$,
e.g.~$\Delta_0^- =1/K^2$.

Performing the sum over the Matsubara frequencies yields
$-2g^2N\sumints\Delta_m = {3\0 2} m^2 $ with
$m^2={g^2NT^2 \0 9}$ and
\beal{lcc9}
\lefteqn{
4\sumint \Delta_m^-\Delta_m P_0K_0 \;=\;
\sumint \lgk {1\0 P_0-\w_p}+{1\0 P_0+\w_p} \rgk
        \lgk {1\0 K_0-\w_k}+{1\0 K_0+\w_k} \rgk }\nn\\
 & = & \int {d^3p\0(2\pi)^3}
      \lek n(\w_p)+n(\w_k) \rek
  \({1\0 Q_0+\w_p+\w_k}-{1\0 Q_0-\w_p-\w_k}\) \nn\\
 & & +\lek n(\w_p)-n(\w_k)\rek
  \({1\0 Q_0+\w_p-\w_k}-{1\0 Q_0-\w_p+\w_k}\)
\eea
with $\w_p^2=p^2+m_\infty^2$. After the frequency sum is
done, we continue $Q_0$ to real values, keeping a small
imaginary part $+i\varepsilon$ in mind.

Since $\w_p-\w_k$ is a soft quantity of order $q$,
\bel{wdiff}
\w_p-\w_k=zq-zq{m_\infty^2\02p^2}-{q^2\02p}(1-z^2)
 +O\({q^4\0p^3}\),
\quad z={\vc p \vc q \0 pq},
\eeq
it is useful to expand
$n(\w_k)=n(\w_p)+n'(\w_p)(\w_k-\w_p)+O(q^2)$. Moreover, the
first two denominators on the r.h.s. of (\ref{lcc9}) are far
from potential zeros for hard loop momenta, where $Q_0$ is
negligible against $\w_p+\w_k$, so one can approximate
$\w_p,\w_k\approx p$ there. The remaining denominators are
those that give rise to collinear singularities at the
light-cone when $m_\infty$ is neglected. However, since
these singularities are only logarithmic, we may also expand
their numerators, dropping the terms that have too little
power at large $p$ to contribute to the $T^2$-part. This
leads to
\beq
\Pi_{00} = 3m^2
  + 2g^2N\int\!{d^3p \0 (2\pi)^3}\,n'(p)\,
   {Q_0^2 \0 Q_0^2 -(\w_p-\w_k)^2}\; .
\label{lc70}
\eeq

Sufficiently far from $Q^2=0$, one can replace $\w_p-\w_k
\to zq$. In the vicinity of the light-cone, however, the
main contribution comes from $|z|\approx 1$ and the second
term in (\ref{wdiff}) becomes important, whereas the
subsequent term is suppressed by $(1-z^2)$. This latter term
is effectively of $O({q^4/p^3})$, because the dominant
contribution is generated when $(1-z^2)\sim m^2/p^2$. We
therefore obtain
\bea
\Pi_{00}(Q) & = & 3m^2
+ 2g^2N\int\!{d^3p\0 (2\pi)^3}\,n'(p)\,
            {Q_0^2\0 Q_0^2+q^2{m_\infty^2\0 p^2} - q^2z^2}
\label{lc90}\\
& = & 3m^2
- {g^2NT^2\0 3}\,
  {3\0\pi^2} \int_0^\infty\!\!d\a{\a^2e^{\a}\0 (e^{\a}-1)^2}\,
  \2\int_{-1}^1\!dz\,{Q_0 \0 \wu{1+{\mu^2\0 \a^2}}Q_0 - qz}
\label{lc100}
\eea
where $\mu^2={m_\infty^2 \0 T^2}=g^2N/6$.

Although the angular integration is no longer decoupled from
the integral over the modulus of the loop momentum, this is
still very close to the structure found in the conventional
hard thermal loop, see~Eq.~(\ref{poohtl}). The difference is
only that $Q_\mu$ does not appear in the scalar product with
a light-like ``unit vector'' which is integrated over, but
is contracted with a timelike one whose zero component is
subject to a certain averaging.

Introducing
\beq
 Y_\mu(\a) = (Y_0(\a),\vc e)
 \with Y_0(\a)= {\wu{\a^2 + \mu^2}\0\a}
\label{ny}
\eeq
and the double averaging
\beq
\lang \phantom{UQ \0 YQ} \rang =
 {3\0 \pi^2}\int_0^\infty\!\!d\a {\a^2 e^\a\0 (e^\a-1)^2}\,
 \1{4\pi}\int\!\! d\Omega\;
 \equiv \int\nolimits_\a \int\nolimits_\Omega
\label{nmit}
\eeq
allows us to write the improved HTL polarization tensor in
the same compact way as the conventional one,
\beq
\Pi_\mn =
 3m^2 \lang U_\mu U_\nu - {UQ \0 YQ}Y_\mu Y_\nu \rang \;.
\label{npi}
\eeq
In this expression $Y_0\not=1$ is essential only in the
denominator, since $\mu\sim g$. Strictly speaking, the lower
bound on the integration variable $\a$ is given by $\L/T$,
but this is negligible when concentrating on the leading
contributions $\sim g^2T^2$.

$\Pi_\mn$ is now defined also for soft lightlike momenta.
The former singularity of $\Pi_{00}$ at $Q^2=0$ has
disappeared and is replaced by $\ln({\mathrm const.}/g)$, to
wit
\bel{poolc}
\Pi_{00}(Q^2=0)  = -{g^2NT^2\0 3}
   \lgk\ln{2\0\mu}+\2-\gamma+{\zeta'(2)\0 \zeta(2)}\rgk\;.
\eeq

There is still a logarithmic branch cut for $|Q_0|<q$. The
imaginary part along this cut, however, is a smooth albeit
nonanalytic function in the momentum variables,
\bel{impoo}
\Im m \Pi_{00}(Q)= {9m^2\0 2\pi}{Q_0\0 q}
\int_{\a_o(Q)}^\infty \!\!d\a{\a^2e^{\a}\0 (e^{\a}-1)^2}
\quad {\rm for} \; Q^2<0
\eeq
and with $\a_o(Q)=\mu Q_o/\wu{-Q^2}$. For $|Q^2|/Q_0^2\gg
g$, the lower integration bound $\a_o(Q)$ is of order $g$,
hence negligible at leading order. However, it becomes of
order 1 when $|Q^2|/Q_0^2\sim g$, and finally goes up to
infinity for $|Q^2|\to 0$, thereby bringing (\ref{impoo})
down to zero.

\subsection{Gluon vertex functions}

For gluon vertex functions more complicated analytic
structures arise. Performing the analogous steps as above,
we find e.g.~for the $000$-component of the 3-gluon-vertex
function
\bea
&&\G_{000}(Q,R,-Q-R)\nn\\
&&=
g^3N\int\!{d^3p\0(2\pi)^3}\,n'(p) \lgk
R_0 \( \1{Q_0+\w_{p-q}-\w_p} \1{R_0+\w_{p-q-r}-\w_{p-q}}
 \right.\right. \nn\\
 &&\quad\quad
+ \left.\1{Q_0-\w_{p-q}+\w_p} \1{R_0-\w_{p-q-r}+\w_{p-q}} \)
 \nn\\
&&\quad
-(Q_0+R_0)\(\1{Q_0+\w_{p-q}-\w_p}\1{Q_0+R_0+\w_{p-q-r}-\w_{p}}
\right. \nn\\
&&\quad\quad \left.\left.
+\1{Q_0-\w_{p-q}+\w_p} \1{Q_0+R_0-\w_{p-q-r}+\w_{p}} \) \rgk
\eea
after dropping all contributions $\ll g^3T^2$.

Without the asymptotic thermal masses there are collinear
singularities when any of the external momenta is
light-like. These are again only logarithmic as long as the
external momenta have different directions. Assuming this,
one can for each denominator in turn expand the energy
differences according to (\ref{wdiff}). This again amounts
to modifying $Q_0\to Q_0(1+{m_\infty^2\02p^2})$ and likewise
$R_0$. After that the two pairs of terms in the round
brackets can be combined by changing $\vc p\to-\vc p$.

As a result, the modified gluon vertex can be cast into the
same form as was possible for the conventional HTL vertex
\beq
  \G_{000} = 3gm^2 \lang \1{(YQ)}
     \lgk {R_0\0(YR)} - {Q_0+R_0\0(Y(Q+R))} \rgk \rang,
\label{eff30}
\eeq
but with the averaging and the vector $Y$ redefined
according to (\ref{nmit}) and (\ref{ny}).

Evaluating the angular integral gives
\beq
\G_{000} = 3gm^2 \int\nolimits_\a \lek
 R_0 \cl M(\bar Q,\bar R)
 -(Q_0+R_0) \cl M(\bar Q,\bar Q+\bar R) \rek
\eeq
where
\bel{tiM}
\bar Q = (Q_0 \wu{1+\mu^2/\a^2},\vc q)
\eeq
and $\cl M$ the Lorentz-invariant function introduced in
Ref.~\cite{FT},
\bel{MFT}
\cl M(K,P)=\1{2\wu{-\Delta}}
 \ln\({KP+\wu{-\Delta}\0KP-\wu{-\Delta}}\)\quad, \quad
\Delta = K^2 P^2-(KP)^2
\eeq

In the conventional HTL result there are logarithmic
singularities at $Q^2=0$, $R^2=0$, and $(Q+R)^2$ as well as
(generally nonsingular) branch points at $\Delta(Q,R)\equiv
\Delta(Q,Q+R)=0$. In the improved result (\ref{eff30}),
these singularities are removed because in the $\a$-average
they are smeared according to (\ref{tiM}).

As concerns higher vertex functions, the above argument that
led to the form (\ref{eff30}) can be essentially repeated.

\subsection{Improved effective action}

Because of the formal similarity of the improved hard
thermal loops to the conventional ones --- we only had to
redefine the 4-vector that is used in the angular average
and extend the averaging --- it is natural to guess that one
can also take over the compact effective action by Braaten
and Pisarski \cite{BPeff}
\beq
\cl S_{\mathrm eff} = -3m^2 {1\0 4} \int\! d^4x F_a^{\mu\a}(x)
 \lang{Y_\a Y_{\beta} \0 (YD)^2_{ab}}\rang F_b^\beta{}_\mu(x)\;,
\label{effBP}
\eeq
but now with the definitions (\ref{ny}) and (\ref{nmit}).
These modifications obviously do not interfere with the
manifest gauge invariance of the HTL effective action.

One can readily verify that (\ref{effBP}) contains the
improved version of the HTL two-point function, but some
manipulations are needed to bring it into the form derived
above (see the Appendix).

In the case of vertex functions, it is advantageous to use
the earlier version of the HTL effective action of Taylor
and Wong \cite{eff}, which is, however, not manifestly gauge
invariant. We therefore demonstrate that the redefinitions
(\ref{ny},\ref{nmit}) do not interfere with its actual gauge
invariance either.

We first write
\beq
 \cl S_{\mathrm eff} = \sum_{n=2}^\infty \cl S_n[A] \; ,
\eeq
where $V_n$ collects the contributions $n$-linear in $A_\mu$.
{}From the improved gluon self-energy we have
\bea
 \cl S_2[A] & = & \2\int{d^4Q\0(2\pi)^4}
   A^{a\mu}(-Q)\,\Pi_\mn(Q)\,A^{a\nu}(Q) \nn\\
 & = & 3m^2\int\!d^4x\Sp
  \lang [H_0(x)]^2 - H(x){Y_0\6_0 \0 Y\6} H(x) \rang
\label{eff10}
\eea
with the abbreviations $H(x)=Y^\mu A_\mu^a(x)\cl T^a$ and
$H_0(x)=Y_0A_0^a(x)\cl T^a$. With the ansatz
\beq
  \cl S = 3m^2\int\!d^4x\Sp\lgk
\lang [H_0(x)]^2 \rang - \lang \phi[H] \rang\rgk \; .
\eeq
one can find a gauge-invariant functional $\phi$ with
``boundary condition'' (\ref{eff10}) as follows.

Under infinitesimal gauge transformations with parameter
$\w$ we have
\beq
\d_\w H_0 = Y_0\6_0 \w - ig[H_0,\w] \quad\mbox{and}\quad
\d_\w H   = Y\6 \w - ig[H,\w] =: \cl D \w \;.
\eeq
Gauge invariance of $\cl S$ implies $\d_\w \cl S[A] = 0$ for
all $\w$. This requires
\beq
2\lang Y_0\6_0 H \rang = \lang \cl D \6_H\phi[H] \rang \;.
\eeq
Up to terms that average to zero, a solution to this
equation is given by a functional satisfying
\bel{phidg}
\6_H\phi[H]= {2\0 \cl D}\, Y_0\6_0 H
\with \1{\cl D} = \1{1-\1{Y\6}[igH,*]}\, \1{Y\6} \;.
\eeq
Note that here there is a small difference to the
conventional case. Because of the new averaging procedure
the inverse of $(Y\6)$ always exists.

Counting only explicit coupling constants (not those
implicit in $Y_0$ which depends on $\mu\propto g$), we have
$g\6_g\phi = \int d^4x\Sp H\6_H\phi$. That is
\beq
g\6_g\phi[H] = 2\int\!d^4x\Sp \sum_{n=0}^\infty
  H\lgk {1\0 Y\6}[igH,*]\rgk^n{1\0 Y\6}Y_0\6_0H \;.
\eeq
Integrating with respect to $g$ yields
\beal{effTW}
 \cl S_{\mathrm eff}[A] & = & 3m^2\int\!d^4x\Sp\lang
 (H_0)^2 + (Y_0\6_0H) F\({1\0 Y\6}[iH,*]\) {1\0 Y\6} H\rang \\
\label{eff20}
  &  & \quad {\rm where} \quad
  F(z)=2\sum_{n=0}^\infty {z^n\0 n+2} \; .\nn
\eea

This functional is gauge invariant by construction. In fact,
it is identical in form with that of Ref.~\cite{eff}, only
the meaning of the symbols has changed. It reproduces the
improved vertex functions in the form in which we had them
obtained in the previous section, since $Y_0$'s in the
numerators can be put to 1. This just drops terms that are
suppressed by powers of $g$, which we have always discarded.

We thus have shown that inclusion of the asymptotic gluon
mass (\ref{mu}) removes the collinear singularities of the
hard thermal loops without spoiling gauge invariance.

\section{Inclusion of fermions}

Defining the two structure functions of the fermion
self-energy $\Sigma$ at finite temperature by
\beq
\Sigma=aQ_0\g_0 + b\vc q\bg \quad,\quad
a=\1{4Q_0}\Sp\g_0\Sigma\qquad
b=-\1{4q^2}\Sp{\vc q\bg}\Sigma\;,
\eeq
the dressed propagator reads
\beq
 S = - { (1+a)Q_0\g_0 - (1+b)\vc q\bg
  \0 (1+a)^2Q_0^2 - (1+b)^2 q^2  }\; .
\label{lc170}
\eeq
The leading high-temperature contributions are
\bel{ab}
a=-{M_f^2\0 2qQ_0}\ln\({Q_0+q\0 Q_0-q}\) \quad , \quad
b={M_f^2\0 q^2}\lgk 1-{Q_0\02q} \ln\({Q_0+q\0 Q_0-q}\)\rgk
\eeq
with $M_f^2= \1{8}g^2C_fT^2$ (the QED case is covered by
replacing $g^2C_f\to e^2$). For soft momenta, this contains
an extra collective mode, the ``plasmino''. However, as with
the plasmon, its residue vanishes exponentially fast with
increasing $q$. For $Q_0,q \gg gT$, the dressed propagator
approaches
\beq
 S(Q^2) = -{Q_\mu\g^\mu + O(M^2/q)
  \0 Q^2 - M_\infty^2 + O(M^4/q^2)} \; ,
\label{Sas}
\eeq
where
\bel{Minf}
 M_\infty^2 = 2(Q_0^2 a+q^2 b) = 2M_f^2.
\eeq
Like the asymptotic gluon mass, this latter result (in
contrast to (\ref{ab})) is even an exact one-loop result
when $Q^2=0$, i.e.~not merely the leading high-temperature
term for $Q_0,q\ll T$ (see the appendix of
Ref.~\cite{Wferm}).

With only external gluons the fermion loop produces the same
HTL contributions as the purely gluonic one-loop diagrams
did, except that instead of $N$ there is an overall factor
of $N_f/2$. For lightlike or nearly lightlike external
momenta, the asymptotic fermion mass $M_\infty$ becomes
important and appears just in place of $m_\infty$.

However, the situation is essentially different when hard
thermal loops with external fermions are considered. The
only hard thermal loops are those with two external fermion
lines. Those are given by a loop which involves necessarily
both internal fermion and gauge boson propagators, which
have different asymptotic masses.

\subsection{Fermion self-energy}

Dressing the hard propagators in the fermion self-energy
with their respective asymptotic masses gives the improved
hard thermal loop
\beal{nSigma}
\Sigma(Q)
&=&-\14 g^2C_f \int{d^3p\0(2\pi)^3}\,[n(p)+\tilde n(p)]
   \(\g_0-{\vc p \bg \0 p} \)
\nn\\
&&\qquad\qquad\times
\lgk \1{Q_0-\w_p+\tilde\w_{p-q}}+\1{Q_0+\w_p-\tilde\w_{p-q}}
\rgk
\eea
where $\tilde n$ is the Fermi-Dirac distribution function and
$\tilde \w_k=\wu{k^2+M_\infty^2}$.

In the difference
\bel{wdifff}
\w_p-\tilde\w_k =
  zq+{m_\infty^2-M_\infty^2\0 2p}-{q^2\02p}(1-z^2)
 +O({q^3\0p^2}),
\quad z={\vc p \vc q \0 pq},
\eeq
the dominant term that shifts the poles in the integrand of
(\ref{nSigma}) is now independent of $Q$ and proportional to
the difference of the two asymptotic masses. The other
correction in (\ref{wdifff}) that is of comparable magnitude
is suppressed for $|z|\approx1$, which makes it effectively
of order $q^3/p^2$.

The two terms in the curly brackets of (\ref{nSigma})
therefore do not combine when exchanging $\vc p\to-\vc p$
and we arrive at an improved HTL fermion self-energy of the
form
\beq
\Sigma(Q)=-M_f^2\int\nolimits_\a^f\int\nolimits_\Omega
\lgk {Y^\mu\g_\mu\0 (YQ)+{dm \0 \a}}
   + {Y^\mu\g_\mu\0 (YQ)-{dm \0 \a}} \rgk
\label{lc200}
\eeq
where
\beq
Y_\mu=Y_\mu(\infty)=(1,\vc e)\, ,\quad
\int\nolimits_\a^f=
 {2\0\pi^2}\int_{0}^\infty \! d\a {\a e^\a\0e^{2\a}-1}\, ,\quad
dm={m_\infty^2-M_\infty^2\0 2 T}\;.
\label{lc210}
\eeq
So the 4-vector $Y$ is light-like as with the conventional
hard thermal loops. The collinear singularity is instead
smeared out by the addition of $\pm dm/\a$ with a slightly
modified averaging prescription for $\a$.

The singularity of $\Re e \Sigma$ at the light-cone is again
cut off by the asymptotic masses with the result
\bea
\Re e\, a(Q^2=0) &=& -{M_f^2\0 2q^2}\lgk
\ln{q\0|dm|}+{5\04}\ln2+1-\gamma+{\zeta'(2)\0 \zeta(2)}\rgk
\\
\Re e\, b(Q^2=0) &=& \Re e\, a(Q^2=0) + {M_f^2\0q^2}
\eea
However, in contrast to the purely gluonic case, the
imaginary part of the fermion self-energy is left unchanged
for all spacelike momenta and remains nonvanishing at the
light-cone,
\beq
\Im m\, a(Q^2=0)=\Im m\, b(Q^2=0)=-{\pi\02}{M_f^2\0q^2}
\eeq
dropping rapidly but smoothly to zero only
for timelike $Q^2\gtrsim g^2 (gT)^2$.

\subsection{Improved effective action with fermions}

In the case of the conventional hard thermal loops, the
effective action for fermionic Green functions is given by
the simplest possibility to bring its bilinear part, which
is determined by the fermion self-energy, into a
gauge-invariant form. With equal ease, this can be done with
the improved HTL fermion self-energy,
\beq
 \cl L_{\rm eff}=-\quer\psi(x)M_f^2
  \int\nolimits_\a^f\int\nolimits_\Omega
  \({Y^\mu\0 iYD+{dm\0\a}}+{Y^\mu\0 iYD-{dm\0\a}}\)
  \g_\mu\psi(x)
\label{lc220}
\eeq
with $D^\mu=\6^\mu{\bf 1}-ig\cl T^a A^{a\mu}$.

This obviously requires that not only the fermion
self-energy but all higher vertex functions can be broken up
in two parts which differ only by the sign in front of $dm$.
This is indeed the case as can be shown by induction.

Let us first explain how the structure of (\ref{lc200})
arises. This simplest of the fermionic Green functions
contains one fermionic and one bosonic propagator inside the
loop. Decomposing each in partial fractions gives rise to a
sum of products of the form
\bel{2pr}
\Delta_F \Delta_B \to \sum_{\s_1\s_2}
       \1{F_0+\s_1 \tilde\w_f}\1{B_0+\s_2 \w_b}
\eeq
where we have denoted the respective momenta by $F_\mu$ and
$B_\mu$ and the $\s_i$ are signs. $F$ and $B$ differ only by
soft momenta, so we choose their orientation such that
$F_\mu\approx+B_\mu$, and keep this convention when we shall
be considering additional propagators in the loop.
Performing now a further partial fractioning on (\ref{2pr})
yields
\bel{2prpf}
\sum_{\s_1\s_2}\1{B_0-F_0+\s_1\w_b-\s_2\tilde\w_f}
\(\1{F_0+\s_1 \tilde\w_f}-\1{B_0+\s_2 \w_b}\) \;.
\eeq
Summing over the Matsubara frequencies will give
$-\s_1\tilde n(\w_f)$ in the first, and $-\s_1 n(\w_b)$ in
the second term. Of the four parts of the sum only two can
contribute to hard thermal loops, namely those where $\tilde
n(\w_f)$ and $n(\w_b)$ acquire the same sign, $\s_1=\s_2$.
Otherwise the contribution will be suppressed by a
hard denominator in the prefactor.

Now let us consider a loop with one additional propagator,
$\Delta_{F'}$ or $\Delta_{B'}$. This brings in another
factor $1/(F_0'+\s'\tilde\w_{f'})$ or $1/(B_0'+\s'\w_{b'})$.
Decomposing each product with the terms of (\ref{2prpf})
into partial fractions, one readily notices that soft
denominators in the prefactors require like signs.
Therefore, we still end up with exactly two contributions to
the hard thermal loop, $\s'=\s_1=\s_2=\pm1$. Moreover, of
all the various combinations only those contribute to hard
thermal loops where propagators of different statistics are
combined so that in the end there is always a sum $\tilde
n+n$ as in (\ref{nSigma}), for the difference in the
arguments of $\tilde n$ and $n$ does not matter.

The same argument applies to higher vertex functions, since
in all the fermionic Green functions there is just enough
power from hard loop momenta to produce a hard thermal loop
$\propto T^2$ when all the energy denominators up to one are
soft.

Usually these two contributions can be identified since only
the first term of the r.h.s. of (\ref{wdifff}) is kept and
this changes sign with the spatial loop integration
variable. For the improved hard thermal loops, however, we
have to keep also the subsequent term in (\ref{wdifff}), which
does not.

Let us finally write down one example for an improved HTL
vertex with external fermions. Either by direct calculation
or by functional differentiation of (\ref{lc220}) one finds for
the quark-quark-gluon vertex
\bea
\lefteqn{\G^\mu(Q,R;R-Q)}\nn\\
\lefteqn{=g\cl T\g_\nu M_f^2\int\nolimits^f_\a\!\!\int\nolimits_\Omega
\lgk {Y^\mu\0 (YQ+{dm\0\a})}{Y^\nu\0 (YR+{dm\0\a})}
    + {Y^\mu\0 (YQ-{dm\0\a})}{Y^\nu\0 (YR-{dm\0\a})} \rgk \;.}
 \label{lc230}
\eea

Just as the analytic structure of the improved fermion
self-energy turned out to be modified in a somewhat
different fashion than was the case for the gluon
self-energy, the vertex functions involve slightly different
functions. Nevertheless, these can again be expressed in
terms of integrals involving the function $\cl M$ introduced
in (\ref{MFT}). For instance,
\bea
\4\Sp\(\g^0\G^0\)&=&
g\cl TM_f^2\int\nolimits_\a^f\lek \cl M(\tilde Q_+,\tilde R_+)
                        +\cl M(\tilde Q_-,\tilde R_-) \rek \\
{\rm with}\quad \tilde Q_+&=&(Q_0+\textstyle{dm\0\a},\vc q)
\quad{\rm and}\quad
\tilde Q_- = (Q_0-\textstyle{dm\0\a},\vc q)\;.
\eea

Again the inclusion of the asymptotic mass (\ref{Minf})
smoothes out the singularities in $\cl M$
in the final integration over $\a$.

\section{Conclusion}

To summarize, dressing the hard propagators of the hard
thermal loops by the asymptotic masses $\propto gT$ that
pertain to the transverse branch of the gluonic excitations
and to the normal branch of the fermionic ones removes all
collinear singularities of the hard thermal loops and also
preserves their gauge invariance. Moreover, the elegant
effective actions of Braaten and Pisarski and of Taylor and
Wong could be generalized to summarize the improved hard
thermal loops in an equally compact form.

The gauge invariance of the improved hard thermal loops
appears to be particularly encouraging to use them in place
of the original hard thermal loops, where the latter lead to
singular results when external lightlike momenta are
involved. However, we have not yet shown that the
systematics of a resummed perturbation theory built on the
now everywhere well-defined hard thermal loops is as it was.
It may well be that the would-be collinear singularities,
which are only logarithmic at the level of hard thermal
loops, build up in higher loop diagrams to a degree that
overpowers the suppression by powers of $g$. Indeed, in
Ref.~\cite{FS} it has been found that in QCD the resummed
one-loop diagrams are sufficiently singular to contribute to
the sublogarithmic terms in eq.~(\ref{poolc}). Actually, the
imaginary part (\ref{impoo}) is even less stable and becomes
large already for timelike momenta with $Q^2/q^2\sim g$,
thus preventing the longitudinal plasmon branch from coming
arbitrarily close to the light-cone. This behaviour, which
is completely opposite to the one observed in the case of
scalar electrodynamics \cite{KRS}, will be the subject of a
separate investigation \cite{FRprep}.

Another place where collinear singularities do not cancel
from a resummed calculation using ordinary hard thermal
loops is the case of real soft photon production \cite{BPS}.
Using the improved hard thermal loops, on the other hand,
gives a finite result when calculating the soft contribution
at resummed one-loop order
\bel{srppr}
E{dW\0d^3p}\bigg|_{\mathrm soft\, contr.}
\simeq {Q_q^2 \a \a_s\02\pi^2}T^2\({M_f\0E}\)^2 \ln\(1\0g\)
\ln\(\L\0M_f\) \;,
\eeq
which coincides with the leading logarithms of
Ref.~\cite{Nie}.

However, there is also a hard contribution which has to
restore independence of the scale $\L$ separating soft from
hard momenta. For this it has to be such that its mass
singularities are also cut off in a way that produces a
$\ln(1/g)$ besides a $\ln(T/\L)$. Just like the internal
hard propagators in the improved hard thermal loops this
requires hard diagrams which are dressed by higher order
diagrams. Again we expect that the main effect will be from
the asymptotic thermal masses, which, as we have mentioned
before, do not depend on a low-momentum limit (but they
depend on $\Lambda$, which would become important in a more
accurate calculation). So the same mechanism that renders
finite the soft contribution should also apply to the hard
one, and the last logarithm in (\ref{srppr}) should combine
into $\ln(T/M_f) \sim \ln(1/g)$.

Clearly, much work is still needed to first verify that an
improved resummed perturbation theory really works as
sketched here, and second to fully evaluate the
sublogarithmic contributions. Encouraged above all by the
gauge invariance of the improved hard thermal loops, we
expect them to play a central role in this and analogous
problems.

\begin{ack}

We are grateful to Hermann Schulz for valuable discussions.

\end{ack}

\section*{Appendix}

In order both to corroborate the results obtained in
sect.~3.1 and to show a potential pitfall in their
derivation, let us recalculate (\ref{lc100}) by writing
first
\bel{pooell}
\Pi_{00}(Q)=-{q^2\0Q^2}\Pi_\ell
 \with \Pi_\ell=B^{\mu\nu}\Pi_{\mu\nu}.
\eeq
(This assumes transversality of $\Pi$, which is guaranteed
by the gauge invariance of the (improved) HTL effective
action.)

The potential HTL contributions of $\Pi_{\mu\nu}$ give
\bea
\Pi_\ell(Q)&=&-2e^2 \sumint \Delta_m \nn\\
&&+4e^2 \sumint \Delta_m\Delta^-_m
 \lek p^2-{(\vc p\vc q)^2\0q^2} +P^2-{(PQ)^2\0Q^2} \rek \;.
\eea
This can be rewritten as
\bel{pil2}
\Pi_\ell(Q)=\underbrace{ 4e^2 \sumint \Delta_m\Delta^-_m
  \lek p^2-{(\vc p\vc q)^2\0q^2} \rek}_{(I)}
+\underbrace{ e^2(4m_\infty^2-Q^2) \sumint
  \Delta_m\Delta^-_m}_{(II)}
\eeq
where, superficially, only the first part appears to be a
hard thermal loop. It yields
\bel{pilI}
\Pi_\ell^{(I)}(Q)=-{Q^2\0q^2}{e^2\02\pi^2}
\int_0^\infty \,dp\,p\,n(p) \int_{-1}^1 dz {z^2-1\0
\( z-Q_0/q\wu{1+m_\infty^2/p^2}\)^2}.
\eeq

The result (\ref{pilI}) has exactly the form one obtains by
expanding the manifestly gauge invariant effective action of
eq.~(\ref{effBP}), which involves one more denominator than the
one eq.~(\ref{effTW}). However, the derivative $n'$ implicit in
the averaging in (\ref{effBP}) is replaced by $-2n$. For
conventional hard thermal loops, this does not make any
difference. But close to the lightcone it does. For example,
the result (\ref{poolc}) is not exactly reproduced--- the $+\2$
is missing.

What goes wrong here is that one has to divide $\Pi_\ell$ by
$Q^2$ in order to obtain $\Pi_{00}$. This way it happens
that, close to the light-cone, the second contribution in
(\ref{pil2}) can no longer be neglected. It reads
\bel{pilII}
\Pi_\ell^{(II)}(Q)=(4m^2_\infty-Q^2){e^2\04\pi^2}
 \int_0^\infty \,dp\,{n(p)\0p}
 {Q^2\0Q_0^2(1+m_\infty^2/p^2)-q^2}.
\eeq
Indeed, this by itself does not give rise to a HTL
contribution, except when the $Q^2$ in the numerator is
removed, which is done in (\ref{pooell}). Then there is a
linear singularity for $Q^2\to0$ that is cut off by
${m_\infty^2/p^2}$ in the denominator, which heaves two
powers of hard momentum into the numerator. This precisely
accounts for the missing contribution to $\Pi_{00}(Q^2=0)$.
Moreover, adding (\ref{pilII}) to (\ref{pilI}) can be shown to
be equivalent to replacing $n$ by $-\2n'$ in the latter so that
the gluon self-energy is exactly as prescribed by the
manifestly gauge-invariant effective action (\ref{effBP}).

\end{document}